\documentclass{aa}
\usepackage{textcomp}
\usepackage{graphicx} %
\usepackage{amsmath} %
\usepackage{amssymb} %
\usepackage{bm} %
\usepackage{upgreek} %
\usepackage{IEEEtrantools} %
\usepackage{multirow}
\usepackage[colorlinks=true,allcolors=blue,urlcolor=blue]{hyperref}
\usepackage{txfonts}
\usepackage[normalem]{ulem} 
\usepackage[dvipsnames]{xcolor}

\newcommand{\eqn}[1]{\text{Eq.~\ref{#1}}}
\newcommand{\sect}[1]{\text{Sect.~\ref{#1}}}
\newcommand{\fig}[1]{\text{Fig.~\ref{#1}}}
\newcommand{\tab}[1]{\text{Table~\ref{#1}}}

\newcommand{\stagger}{\textsc{stagger}}
\newcommand{\balder}{\textsc{balder}}
\newcommand{\multitd}{\textsc{multi3d}}
\newcommand{\cobold}{\textsc{co$^{5}$bold}}
\newcommand{\blue}{\textsc{blue}}
\newcommand{\mtd}{\textlangle3D\textrangle}
\newcommand{\marcs}{\textsc{marcs}}
\newcommand{\atmo}{\textsc{atmo}}

\newcommand{\kms}{\mathrm{km\,s^{-1}}}

\newcommand{\lggf}{\log{gf}}

\newcommand{\lgeps}[1]{\log{\epsilon_{\mathrm{#1}}}}
\newcommand{\sh}{S_{\mathrm{H}}}
\newcommand{\lgr}{\log{\tau_{\mathrm{R}}}}
\newcommand{\dex}{\mathrm{dex}}

\newcommand{\nm}{\mathrm{nm}}

\newcommand{\eV}{\mathrm{eV}}
\newcommand{\vmic}{\xi_{\text{mic}}}
\newcommand{\vmac}{\xi_{\text{mac}}}

\begin{document} 

\title{The 3D non-LTE solar nitrogen abundance from atomic lines}
\author{A.~M.~Amarsi\inst{1}
\and
N.~Grevesse\inst{2,3}
\and
J.~Grumer\inst{1}
\and
M.~Asplund\inst{4}
\and
P.~S.~Barklem\inst{1}
\and
R.~Collet\inst{5}}
\institute{Theoretical Astrophysics, Department of Physics and Astronomy, 
Uppsala University, Box 516, SE-751 20 Uppsala, Sweden\\
\email{anish.amarsi@physics.uu.se}
\and
Centre Spatial de Li\`ege, Universit\'e de Li\'ege, avenue Pr\'e Aily, 
B-4031 Angleur-Li\`ege, Belgium
\and
Space sciences, Technologies and Astrophysics Research (STAR)
Institute, 
Universit\'e de Li\`ege, All\'ee du 6 ao\^ut, 17, B5C, 
B-4000 Li\`ege, Belgium
\and
Research School of Astronomy and Astrophysics, Australian National
University, Canberra, ACT 2600, Australia
\and
Stellar Astrophysics Centre, Department of Physics and Astronomy, Aarhus
University, Ny Munkegade 120, DK-8000 Aarhus C, Denmark}

\abstract{Nitrogen is an important element
in various fields of stellar and Galactic astronomy,
and the solar nitrogen abundance is crucial as a yardstick
for comparing different objects in the cosmos.
In order to obtain a precise and accurate value for this abundance,
we carried out \ion{N}{I} line formation calculations in
a 3D radiative-hydrodynamic \stagger{} model 
solar atmosphere, in full 3D non-local thermodynamic equilibrium (non-LTE),
using a model atom
that includes physically-motivated descriptions for the inelastic
collisions of \ion{N}{I} with free electrons and with neutral hydrogen.
We selected five \ion{N}{I} lines of high excitation energy
to study in detail, based on their strengths and on their being relatively 
free of blends. 
We found that these lines are slightly strengthened
from non-LTE photon losses and from 3D granulation effects,
resulting in negative abundance corrections of around
$-0.01\,\dex$ and $-0.04\,\dex$ respectively.
Our advocated solar nitrogen abundance is $\lgeps{N}=7.77$,
with the systematic $1\sigma$ uncertainty estimated to be 
$0.05\,\dex$. 
This result is consistent with earlier studies after
correcting for differences in line selections
and equivalent widths.}

\keywords{atomic processes --- radiative transfer --- line: formation --- 
Sun: abundances --- Sun: photosphere --- stars: abundances}

\date{Received 5 March 2020 / Accepted 27 March 2020}

\maketitle

\section{Introduction}
\label{introduction}

Nitrogen is an important element in many different areas of astrophysics.
Its dominant isotope, $^{14}$N, is synthesised via the CNO-cycle:
its main origins in the universe are thought to be Asymptotic Giant Branch
stars and, at least at low metallicities, fast-rotating massive stars 
\citep[e.g.][]{2014PASA...31...30K,2016MNRAS.458.3466V}.
Nitrogen abundances shed light on the structure and evolution
of stars, and of galaxies, via measurement in 
a variety of different cosmic objects including
measured in hot and cool stars
in our Galaxy and its satellites
\citep[e.g.][]{2017MNRAS.465..501S,
2018A&A...618A.102M,2019MNRAS.489.1533L,2019MNRAS.484.3093S},
in interstellar media
\citep[e.g.][]{2017MNRAS.469..151B,2018MNRAS.478.2315E},
and in damped Lyman-$\upalpha$ systems
\citep[e.g.][]{2014MNRAS.444..744Z,2019ApJ...874...93B}.

The solar abundance of nitrogen
is then crucial for having
a solid yardstick with which to compare the 
different cosmic objects discussed above.
Nitrogen forms highly volatile compounds that do not efficiently
condense into grains; thus its abundance
as measured from meteorites does not reflect the 
initial abundance of the solar system at the time of its birth
\citep{2019arXiv191200844L}.  As such,
the most reliable way to infer 
the initial solar system abundance of nitrogen is 
through spectroscopy of the solar photosphere.

The challenge is to derive the solar abundance with both
high precision and high accuracy. 
In the current standard set of solar abundances
\citep{2009ARA&amp;A..47..481A}, nitrogen has
$\lgeps{N}=7.83\pm0.05$\footnote{$\lgeps{A}\equiv\log_{10}
\left(\frac{N_{\mathrm{A}}}{N_{\mathrm{H}}}\right)+12$,
where $N_{\mathrm{A}}$ and $N_{\mathrm{H}}$ 
are the number densities
of nuclei of element $\mathrm{A}$ and of hydrogen}.
This value is primarily based on 
optical and near-infrared \ion{N}{I} lines
($\lgeps{N}=7.78\pm0.04$),
and near-infrared rotational-vibrational ($\Delta\nu=1$) NH lines
($\lgeps{N}=7.88\pm0.03$); secondary diagnostics
include the near-infrared pure-rotational ($\Delta\nu=0$) NH lines
as well as various CN electronic bands.
For their analysis, the authors employed 
a three dimensional (3D) radiative-hydrodynamic model of the solar 
atmosphere, calculated using the \stagger{} code \citep{Nordlund:1995,
2018MNRAS.475.3369C}; this particular model was 
discussed in, for example, Section 3 of \citet{2015A&amp;A...573A..25S}.

There have been
surprisingly few other recent studies of the solar nitrogen abundance
presented in the literature, at least after considering 
the broad importance of this element in astronomy.
\citet{2009A&A...498..877C} presented an analysis of
the \ion{N}{I} lines using an independent 3D model,
calculated using \cobold{} \citep{2012JCoPh.231..919F}.
Their advocated value is $\lgeps{N}=7.86\pm0.12$, 
which is $0.08\,\dex$ higher than the \ion{N}{I} value from 
\citet{2009ARA&amp;A..47..481A}, but nevertheless consistent
within the stipulated $1\sigma$ uncertainties.

A potential shortcoming of these two previous studies
lies in their treatment of
departures from local thermodynamic equilibrium (LTE).
There is a general consensus that
\ion{N}{I} is susceptible to departures from LTE,
that may amount to around $0.05\,\dex$ in 
solar type stars \citep{2005PASJ...57...65T}.
\citet{2009ARA&amp;A..47..481A} applied non-LTE abundance
corrections from \citet{1996A&A...305..275R}
based on a one-dimensional (1D) model solar,
while \citet{2009A&A...498..877C} calculated non-LTE abundance
corrections on their temporally- and horizontally-averaged 3D model.

There are two important developments that make
revisiting the nitrogen abundance inferred from 
\ion{N}{I} lines of importance. 
First, it is now possible to carry out full 3D non-LTE radiative
transfer for spectrum synthesis and abundance analyses,
without making significant compromises on the 
resolution of the adopted model solar atmosphere, nor on
the complexity of the non-LTE model atom.
In this approach, the departures from LTE are calculated
consistently with the 3D model solar atmosphere itself,
taking into full consideration non-vertical radiative 
transfer.

Secondly, non-LTE model atoms can now be constructed
that are much more reliable than ever before.
This is in large part thanks to the rapid progress made recently in
modelling low-energy inelastic collisions, with both free electrons,
and neutral hydrogen. A poor description of
such processes can often be the main limitation in
non-LTE models \citep{2016A&amp;ARv..24....9B}.
In the past, classical or semi-empirical approaches were 
commonly used to model
them, such as that of \citet{1962ApJ...136..906V} 
for electron collisions, and the Drawin recipe 
\citep{1969ZPhy..225..483D,1968ZPhy..211..404D}
as presented in
\citet{1984A&amp;A...130..319S} and 
\citet{1993PhST...47..186L} for hydrogen collisions.
These are of limited validity however;
the latter being so uncertain, a
fudge factor $\sh$ is usually applied
and calibrated so as to match the observations.
For nitrogen, the situation today is more positive:
ab initio cross-sections for the electron collisions
are available via the $B$-spline $R$-matrix (BSR) method
\citep{2014PhRvA..89f2714W}; cross-sections for the hydrogen collisions 
involving low-lying levels have been calculated via the 
Landau-Zener model coupled with
a linear combination of atomic orbitals (LCAO) method
\citep{2019A&A...625A..78A}, and for more highly-excited 
levels
they can be estimated using the free electron method
\citep{1985JPhB...18L.167K,kaulakys1986free,1991JPhB...24L.127K},
which is valid in the Rydberg regime.

Our aim here is to revisit the solar nitrogen abundance
using the latest spectrum synthesis methods.
Similar to our recent studies of
oxygen \citep{2018A&A...616A..89A} and carbon \citep{2019A&A...624A.111A},
we use the 3D non-LTE radiative transfer code
\balder{} (our modified version of \multitd{}),
a 3D \stagger{} model solar atmosphere, and a new
non-LTE model atom that is based on the BSR and LCAO/free electron 
methods for the inelastic collisions.
We demonstrated that this approach
successfully reproduce the centre-to-limb variations of 
the \ion{O}{I} $777\,\nm$ lines, as well as various
lines of \ion{C}{I}, without having to resort to 
any calibrated fudge factors that would always
be necessary with the Drawin recipe and with 1D models.

The rest of this paper is structured as follows.
We review the method in \sect{method}.
We discuss the nature of the 3D non-LTE effects 
on \ion{N}{I} in
\sect{results}.
We present and discuss our 
advocated solar nitrogen abundance in
\sect{discussion}, and compare the result
with those from previous studies.
We summarise and present some concluding remarks
in \sect{conclusion}.

\section{Method}
\label{method}

\begin{table*}
\begin{center}
\caption{Parameters of the five \ion{N}{I} lines used in the abundance analysis: wavelengths in air $\lambda_{air}$; electronic configurations and spectroscopic terms of the lower and upper states, with excitation energies $\chi_{\text{low}}$ and $\chi_{\text{up}}$; oscillator strengths $f$; natural broadening coefficients $\gamma_{\text{rad.}}$ (FWHM in angular frequency units); and ABO broadening cross-sections $\sigma_{\mathrm{H}}$ at reference velocity $\varv=10^{4}\,\mathrm{m\,s^{-1}}$ and exponents $\alpha_{\mathrm{H}}$, such that the cross-section is proportional to $\varv^{-\alpha}$ \citep{1995MNRAS.276..859A}. Data sources are the same as used to construct the non-LTE model atom (\sect{methodatom}); in particular, the transition probabilities were drawn from \citet{2002A&A...385..716T}.}
\label{tab:n1linelist}
\begin{tabular}{c | c c c c c c c c c}
\hline
\multicolumn{1}{c|}{Line label} &
\multicolumn{1}{c}{$\lambda_{\text{air}} / \mathrm{nm}$} &
\multicolumn{1}{c}{Lower} &
\multicolumn{1}{c}{$\chi_{\text{low}} / \mathrm{eV}$} &
\multicolumn{1}{c}{Upper} &
\multicolumn{1}{c}{$\chi_{\text{up}} / \mathrm{eV}$} &
\multicolumn{1}{c}{$\lggf$} &
\multicolumn{1}{c}{$\log(\gamma_{\text{rad.}} / \mathrm{s^{-1}})$} &
\multicolumn{1}{c}{$\sigma_{\mathrm{H}} / a_{0}^{2}$} &
\multicolumn{1}{c}{$\alpha_{\mathrm{H}}$ }\\
\hline
\hline
\multicolumn{1}{l|}{\ion{N}{I}          744} &
$   744.229$ &
$\mathrm{3s\,^{4}P_{3/2}}$ &
$    10.330$ &
$\mathrm{3p\,^{4}S^{o}_{3/2}}$ &
$    11.996$ &
$    -0.403$ &
$+     8.751$ &
$         530$ &
$    0.2271$ \\
\multicolumn{1}{l|}{\ion{N}{I}          822} &
$   821.633$ &
$\mathrm{3s\,^{4}P_{5/2}}$ &
$    10.336$ &
$\mathrm{3p\,^{4}P^{o}_{5/2}}$ &
$    11.844$ &
$+     0.138$ &
$+     8.745$ &
$         497$ &
$    0.2286$ \\
\multicolumn{1}{l|}{\ion{N}{I}          863} &
$   862.923$ &
$\mathrm{3s\,^{2}P_{3/2}}$ &
$    10.690$ &
$\mathrm{3p\,^{2}P^{o}_{3/2}}$ &
$    12.126$ &
$+     0.077$ &
$+     8.713$ &
$         576$ &
$    0.2337$ \\
\multicolumn{1}{l|}{\ion{N}{I}          868} &
$   868.340$ &
$\mathrm{3s\,^{4}P_{3/2}}$ &
$    10.330$ &
$\mathrm{3p\,^{4}D^{o}_{5/2}}$ &
$    11.758$ &
$+     0.106$ &
$+     8.740$ &
$         481$ &
$    0.2306$ \\
\multicolumn{1}{l|}{\ion{N}{I}         1011} &
$   1010.89$ &
$\mathrm{3p\,^{4}D^{o}_{3/2}}$ &
$    11.753$ &
$\mathrm{3d\,^{4}F_{5/2}}$ &
$    12.979$ &
$+     0.444$ &
$+     7.773$ &
$         750$ &
$    0.2671$ \\
\hline
\end{tabular}
\end{center}
\end{table*}

\begin{table}
\begin{center}
\caption{Equivalent widths of the \ion{N}{I} features, measured at solar disk centre. The blending contributions from CN lines are also shown, as are the contribution to the equivalent widths from the \ion{N}{I} lines themselves.}
\label{tab:n1widths}
\begin{tabular}{c | c c c}
\hline
\multicolumn{1}{c|}{Line label} &
\multicolumn{1}{c}{$W_{\text{total}} / \mathrm{pm}$} &
\multicolumn{1}{c}{$W_{\text{blend}} / \mathrm{pm}$} &
\multicolumn{1}{c}{$W_{\text{\ion{N}{I}}} / \mathrm{pm}$} \\
\hline
\hline
\multicolumn{1}{l|}{\ion{N}{I} 744} &
$0.310$ &
$0.075$ &
$0.235$ \\
\multicolumn{1}{l|}{\ion{N}{I} 822} &
$0.770$ &
$\text{---}$ &
$0.770$ \\
\multicolumn{1}{l|}{\ion{N}{I} 863} &
$0.620$ &
$0.210$ &
$0.410$ \\
\multicolumn{1}{l|}{\ion{N}{I} 868} &
$0.865$ &
$0.115$ &
$0.750$ \\
\multicolumn{1}{l|}{\ion{N}{I} 1011} &
$0.275$ &
$0.075$ &
$0.200$ \\
\hline
\end{tabular}
\end{center}
\end{table}

\begin{figure}
    \begin{center}
        \includegraphics[scale=0.31]{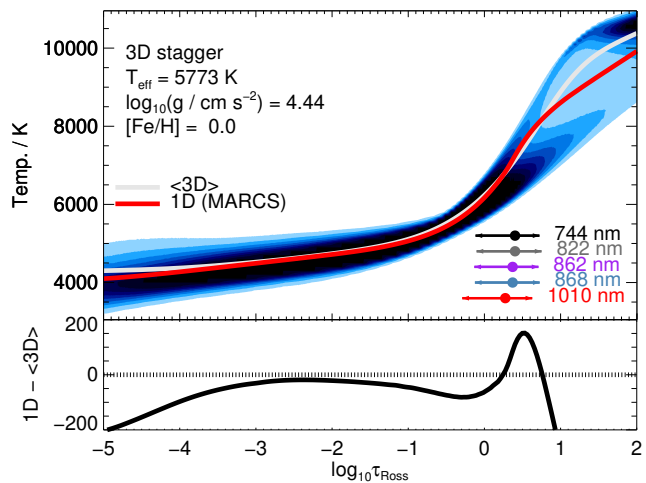}
        \caption{Temperatures as functions of Rosseland mean 
        optical depth for the different model solar atmospheres
        considered in this study (top),
        and residual difference between the 1D and \mtd{} models
        (bottom).
        The optical depths corresponding to peak line formation as well
        as the interquartile ranges of the $5$ \ion{N}{I} lines
        are indicated; these values
        are based on the line-integrated
        contribution function to the line 
        depression in the disk-centre intensity 
        (Eq. 15 of \citealt{2015MNRAS.452.1612A}),
        calculated using the \mtd{} model solar atmosphere.}
        \label{fig:atmos}
    \end{center}
\end{figure}

\begin{figure*}
    \begin{center}
        \includegraphics[scale=0.31]{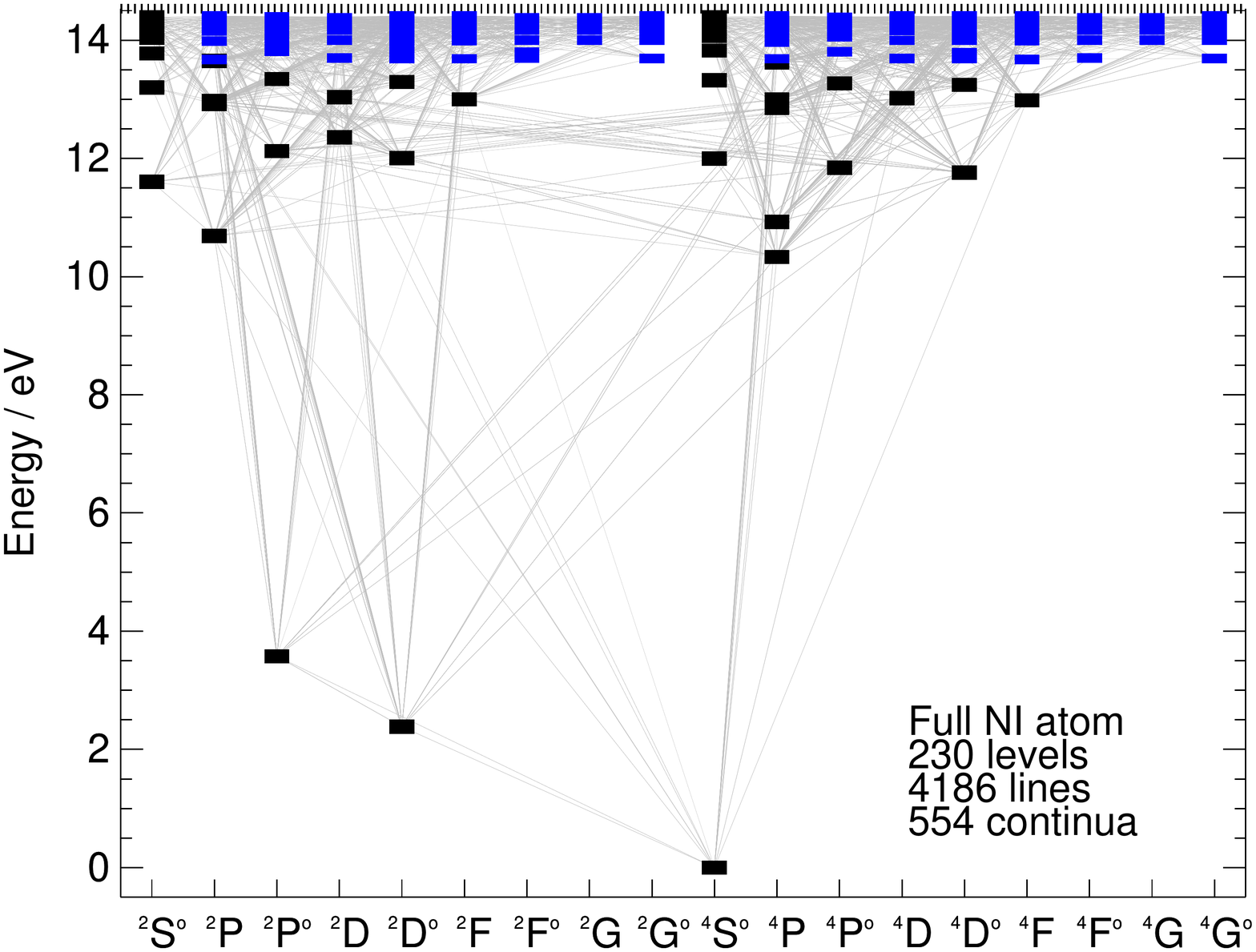}
        \includegraphics[scale=0.31]{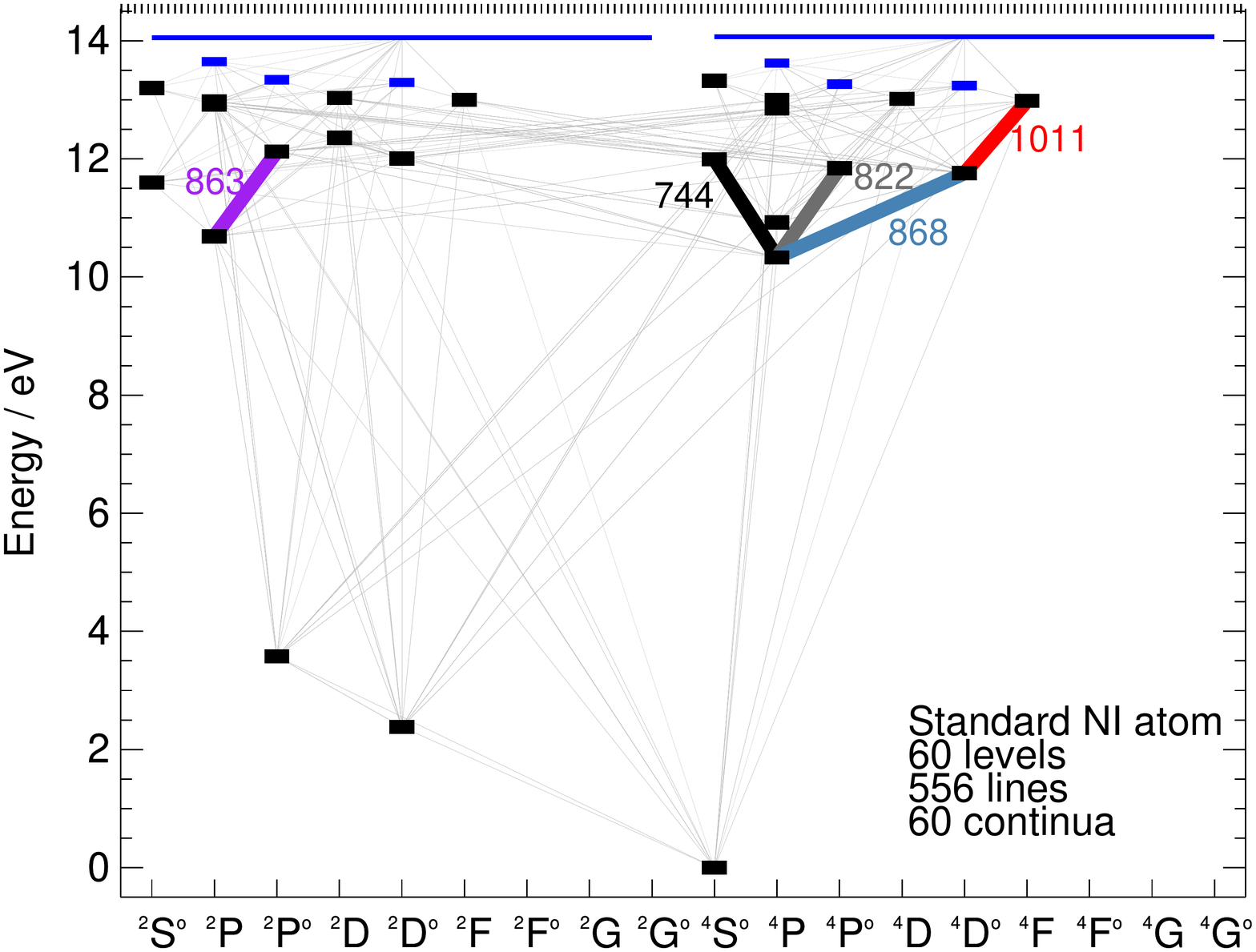}
        \caption{Grotrian diagrams for \ion{N}{I} in the comprehensive
        starting non-LTE model atom (left), and in the resulting 
        non-LTE model atom
        after reducing its complexity (right).
        Energy levels are indicated as short black horizontal lines; 
        those levels for which
        fine structure is not resolved are indicated as 
        short blue horizontal lines, and
        super levels are shown as long blue horizontal lines.
        All bound-bound radiative transitions considered in the non-LTE
        iterations are shown in grey. The five abundance indicators
        are labelled and marked specially in the right panel.}
        \label{fig:grotrian}
    \end{center}
\end{figure*}

\subsection{Abundance indicators and observational data}
\label{methoddiagnostics}

There are more than $20$ different \ion{N}{I} lines
in the solar spectrum that have been used as abundance
indicators in the past \citep{1990A&amp;A...232..225G,
1990asos.conf...59B,2009A&A...498..877C}.
The lines are all of rather high excitation energy
($\chi_{\text{low}}\gtrsim10.3\,\eV$),
forming deep in the solar atmosphere
as we show in \fig{fig:atmos}
($-0.2\lesssim\lgr\lesssim0.7$, at solar disk centre).
All tend to be very weak, and suffering from blends or from
perturbations from nearby lines.
After carefully considering their shapes in the solar atlas, 
we selected just five \ion{N}{I} lines
that are strong enough and sufficiently clean 
to serve as reliable abundance indicators.

We list the parameters of the 
five abundance indicators in \tab{tab:n1linelist}.
Their oscillator strengths were drawn from \citet{2002A&A...385..716T},
which were calculated using the multiconfigurational
Hartree-Fock (MCHF) method, 
including relativistic effects via the Breit-Pauli
Hamiltonian to represent LS-term mixing 
\citep{Fischer_book_1997,2016JPhB...49r2004F}.
Their calculations were
performed with an early version of the MCHF Atomic Structure Package,
\textsc{Atsp2k} 
\citep{2000CoPhC.128..635F,2007CoPhC.176..559F}.
As a sanity check, we carried out our own
fully-relativistic multiconfigurational Dirac-Hartree-Fock 
(MCDHF) calculations, using the latest version of the \textsc{Grasp} 
code \citep{2019CoPhC.237..184F},
that is available as open-source\footnote{The CompAS \textsc{Grasp} repository:
\href{https://github.com/compas/grasp}{github.com/compas/grasp}}.
The agreement between these two approaches was found to be
better than $0.01\,\dex$.

We present the measured equivalent widths of
the \ion{N}{I} features, the CN blends,
and thus the \ion{N}{I} lines, in \tab{tab:n1widths}.
Even with our stringent selection, four of the five selected \ion{N}{I} 
lines are significantly blended by CN lines of high excitation energy.
Although these CN lines were detected via a 3D LTE synthesis,
the CN contribution to the equivalent widths of the 
\ion{N}{I} features were estimated empirically 
by measuring the strengths of neighbouring CN lines 
in the same band.  This approach is preferred over
using the 3D LTE synthesis to estimate the blending contribution,
because it is less sensitive to modelling errors:
namely, neglected non-LTE effects and
deficiencies in the 3D model atmosphere
in the cool upper regions where the molecules form,
$\lgr\approx-1.5$; as well as systematic errors 
in the absolute $\lggf$ values of the CN lines.
Since the CN blends form much higher in the atmosphere
compared to the \ion{N}{I} lines,
their contribution to the total
equivalent width can therefore simply be subtracted,
to give the \ion{N}{I} equivalent widths.

Although the equivalent width of the \ion{N}{I} $822\,\nm$ line 
is not significantly affected by CN blends, the wings of this 
line are visibly perturbed in the solar spectrum.
Consequently, the five \ion{N}{I} lines
are given equal weight in the
abundance analysis presented below.

To minimise the impact of blends, the analysis is based
on disk-centre intensities rather than on disk-integrated fluxes.
The equivalent widths were measured in both
the Jungfraujoch \citep{1973apds.book.....D} and Kitt Peak 
\citep{1984SoPh...90..205N} atlases, both data sets having
extremely high signal-to-noise ratios and spectral resolutions
\citep{1995ASPC...81...32D,1999SoPh..184..421N}.
The two measurements of 
$W_{\ion{N}{I}}$ agree to better than $5\%$, and the final values in 
\tab{tab:n1widths} are the average of the two atlases.

\subsection{Line formation calculations}
\label{methodbalder}

The 3D non-LTE radiative transfer code \balder{},
our modified version of \multitd{} 
\citep{2009ASPC..415...87L_short}, was 
used to calculate synthetic
solar spectra to compare to the observational data.
More details about the code can be found in our previous
papers \citep[e.g.][]{2018A&A...615A.139A}.
In short, \balder{} solves the radiative transfer equation 
simultaneously with
the equations of statistical equilibrium,
following the single-transition preconditioning
scheme described in Sect. 2.4 of
\citet{1992A&amp;A...262..209R}.
During the non-LTE iterations, 
an integral solver is used to solve the radiative transfer
equation on short characteristics \citep{2013A&amp;A...549A.126I}.
Once the populations have converged, a final calculation
is carried out on long characteristics.
The true continuum intensity is also calculated at this point,
by carrying out another calculation with all line opacities set to
zero.  The equation-of-state (EOS)
and background line and continuous opacities are
calculated using the code \blue{}
(Sect. 2.1.2 of \citealt{2016MNRAS.463.1518A}).

The two main inputs for \balder{} are the model
atmosphere, and the non-LTE model atom.
We discuss these in turn in \sect{methodatmosphere}
and \sect{methodatom}, below.
For a given model atmosphere and a given non-LTE model atom,
radiative transfer calculations were performed independently for
different nitrogen abundances, in steps
of $0.2\,\dex$ around a central value of $\lgeps{N}=7.83$.

\subsection{Model solar atmospheres}
\label{methodatmosphere}

We illustrate the different model solar atmospheres
used in this work in \fig{fig:atmos}.
The main results of this study are based on
a 3D radiative-hydrodynamic simulation
that was calculated using the \stagger{} code \citep{Nordlund:1995,
2018MNRAS.475.3369C}. 
This model has a mean effective temperature of 
$5773\,\mathrm{K}$, and was constructed using the solar abundance
set of \citet{2009ARA&amp;A..47..481A};
this abundance set was also adopted when
calculating the EOS and background opacities for the 
line formation calculations.
More details about this 3D model
can be found in our earlier studies of 3D non-LTE line formation in the Sun
\citep{2018A&A...616A..89A,2019A&A...624A.111A}.
The detailed 3D non-LTE radiative transfer was calculated on
eight snapshots of this model, equally spaced over $21\,\text{hours}$ of
solar time.

In addition, calculations were performed on two different 1D 
hydrostatic model solar atmospheres,
to allow for a differential study of the 3D effects
and also to aid future comparisons of this work.
First, calculations were performed
on the temporally- and horizontally-average of 
the 3D model solar atmosphere.
Details of its construction can be found in
\citet{2018A&A...616A..89A}.
The solar abundance set of \citet{2009ARA&amp;A..47..481A}
was adopted when calculating the EOS and background opacities
for the line formation calculations.
We refer to this as the \mtd{} model hereafter.

Secondly, calculations were performed on
the standard 1D \marcs{} model solar atmosphere
\citep{2008A&amp;A...486..951G}.
This model was constructed using
the solar abundance set of \citet{2007coma.book..105G};
as such, we took care to adopt this same abundance set
when calculating the EOS and background opacities
for the line formation calculations.
We verified that the \marcs{} model
gives consistent results (to better than $0.01\,\dex$ in
terms of inferred abundances) with our 
\atmo{} model \citep[][Appendix A]{2013A&amp;A...557A..26M},
the latter being constructed with an identical equation of state and
radiative transfer solver as used for the 3D \stagger{} model. 
This is not too surprising, given that both the \marcs{}
and the \atmo{} models were constructed with the same
underlying physical assumptions (1D, hydrostatic, LTE)
with identical mixing-length descriptions 
\citep{1958ZA.....46..108B,1965ApJ...142..841H}
and very similar background opacity data sources
as well as EOS.
We refer to the \marcs{} model
as the 1D model hereafter.

In general it is necessary to include two tuneable
line broadening parameters
in spectral line synthesis calculations based on 1D model atmospheres
\citep[e.g.][Chapter 17]{2008oasp.book.....G}:
microturbulence $\vmic$, and macroturbulence $\vmac$.
These roughly account for line broadening due to velocity gradients
and temperature inhomogeneities associated with stellar granulation,
on scales much shorter than and much larger than one optical path length,
respectively.  
For the 1D calculations with \balder{},
a depth-independent microturbulence of $\vmic=1\,\kms$ was
adopted; the weak nitrogen lines are not particularly sensitive
to this parameter, and 
increasing $\vmic$ even to $2\,\kms$ only changed the
results by $0.01\,\dex$ for the strongest lines in the line selection
(the \ion{N}{I} $822$ and $868\,\nm$).
Macroturbulent broadening as defined above conserves equivalent widths,
as such the choice of $\vmac$ is inconsequential to the analysis
presented here.
Radiative transfer calculations on 3D model atmospheres
naturally take into account these broadening effects
without having to include extra tuneable parameters
\citep{2000A&amp;A...359..729A}.

\subsection{Non-LTE model atom}
\label{methodatom}

A new non-LTE model atom for \ion{N}{I} was constructed for this study.
The method of construction follows that presented
in our previous studies of \ion{O}{I}
\citep{2018A&A...616A..89A}
and \ion{C}{I} \citep{2019A&A...624A.111A};
we present an overview of the different ingredients in the model here,
and refer the reader to those papers for further details.

We illustrate the ``full'' and ``standard''
non-LTE model atoms in \fig{fig:grotrian}.
As in our previous studies, the standard non-LTE model
atom that was used for the production runs was generated 
by first constructing a full model,
in which we tried to include a complete description
of \ion{N}{I}, without attempting to minimise
the overall complexity of the model.
The complexity of the full model was then reduced
by averaging together certain levels and transitions,
so as to reduce the computational cost
of the 3D non-LTE calculations,
while retaining the most relevant physics in the model
and thus not compromising the accuracy of the final results.
It is important to note that prior to carrying out
the final solution of the radiative transfer equation,
the departure coefficients from the standard model
were applied to the LTE populations of the full model.
This was done to ensure that the energy levels,
oscillator strengths, partition functions, and so on
were as accurate as possible for the calculation
of the synthetic solar spectra.

The full model consists of 230 levels of \ion{N}{I}
along with the low-lying $\mathrm{2p^{2}\,^{3}P_{0,1,2}}$ and
$\mathrm{2p^{2}\,^{1}D_{2}}$ levels of
\ion{N}{II}. The primary data source was
the NIST Atomic Spectra Database \citep{NIST_ASD}:
LSJ energies and oscillator strengths were taken
from here, the original data coming from
\citet{gallagher1993tables}, and from
\citet{1989PhRvA..40.3721Z}, \citet{1991JPhB...24..933B},
\citet{1991A&AS...88..505H}, \citet{1995PhRvA..51.3588M},
\citet{2002A&A...385..716T}, and the
unpublished 1994 Opacity Project data set of Burke and Lennon.
Missing atomic data were then included, taking theoretical LS energies,
oscillator strengths, photoionisation cross-sections,
and natural broadening coefficients from The Opacity Project Database
\citep[TOPbase;][]{1993A&amp;A...275L...5C}.
Broadening coefficients
via elastic neutral hydrogen collisions were obtained by
interpolating the tables of Anstee, Barklem, and O'Mara (ABO;
\citealt{1995MNRAS.276..859A},
\citealt{1997MNRAS.290..102B}, and \citealt{1998MNRAS.296.1057B}).

The cross-sections for excitation and ionisation of 
\ion{N}{I} via electron collisions
\phantomsection\begin{IEEEeqnarray}{rCl}
\label{eq:electron1}
    \mathrm{N+e^{-}}&\leftrightarrow& \mathrm{N^{*}+e^{-}} \\
\label{eq:electron2}
    \mathrm{N+e^{-}}&\leftrightarrow& \mathrm{N^{+}+2e^{-}}
\end{IEEEeqnarray}
were taken from
\citet{2014PhRvA..89f2714W}, which are based on
the BSR method \citep{2006CoPhC.174..273Z}.
These data are complete up to 
$\mathrm{2p^{2}\,3d\,^{2}D}$. For higher
levels, for electron excitation the semi-empirical
recipe of \citet{1962ApJ...136..906V} was used
(this recipe is based on the permitted radiative transition
probability; 
for missing permitted lines, 
and for forbidden lines,
flat dimensionless collision strengths
of $\Upsilon=1.0$, and of 
$\Upsilon=0.1$, were respectively assumed instead);
while for electron ionisation the empirical recipe of 
\citet{1973asqu.book.....A} was used.

The cross-sections for excitation and charge transfer of
\ion{N}{I} via hydrogen collisions
\phantomsection\begin{IEEEeqnarray}{rCl}
\label{eq:hydrogen1}
    \mathrm{N+H}&\leftrightarrow& \mathrm{N^{*}+H} \\
\label{eq:hydrogen2}
    \mathrm{N+H}&\leftrightarrow& \mathrm{N^{+}+H^{-}}
\end{IEEEeqnarray}
were taken from \citep{2019A&A...625A..78A},
which were calculated using the Landau-Zener model 
for non-adiabatic transition probabilities combined 
with a two-electron LCAO method for the electronic structure
(\citealt{2016PhRvA..93d2705B}; see also 
the asymptotic method of \citealt{2013PhRvA..88e2704B}).
As motivated in \citet{2018A&A...616A..89A,2019A&A...624A.111A},
we added to the LCAO cross-sections, the cross-sections
calculated using the \citet{2017ascl.soft01005B} code
that implements the free electron method
\citep{1985JPhB...18L.167K,kaulakys1986free,1991JPhB...24L.127K},
and redistributing the data to account for different spin states
(see Eqs 8 and 9 of \citealt{2016A&amp;ARv..24....9B}).
Finally, the cross-sections for ionisation of 
\ion{N}{I} via hydrogen collisions
\phantomsection\begin{IEEEeqnarray}{rCl}
\label{eq:hydrogen3}
    \mathrm{N+H}&\leftrightarrow& \mathrm{N^{+}+H+e^{-}}
\end{IEEEeqnarray}
were calculated using Eq. 8 of \citet{1985JPhB...18L.167K}.

The TOPbase oscillator strengths and the collisional 
cross-sections adopted here were all calculated under LS-coupling;
that is, without resolving fine structure.
However, the non-LTE model atom does resolve fine structure.
For consistency, the LS oscillator strengths and collisional
cross-sections were redistributed onto the LSJ levels.
The TOBbase lines connecting two LSJ levels from NIST
were redistributed using the tables in
Sect. 27 of \citet{1973asqu.book.....A};
this is valid under the assumption of pure LS coupling.

The collisions were redistributed to account for fine structure
in the same way that was described in
Sect. 2.4.1 of \citet{2018A&A...616A..89A}.
In summary, the rate coefficient from one LS level $x$, to another
LS level $y$, was redistributed 
among LSJ sublevels by dividing
the LS rate coefficient in a given direction 
$x\rightarrow y$ by the total number of 
LSJ target sublevels ($y_{1}$, $y_{2}$, \ldots).
This preserves the total LS rate per perturber per unit time
into the LS level $y$, in the limit where
the energy splitting due to fine structure goes to zero.
The rate coefficient in the reverse direction
$y\rightarrow x$ was then calculated 
strictly using the principle of detailed balance.
Finally, collisions within fine structure sublevels
($y_{1}\leftrightarrow y_{2}$, \ldots)
were introduced and given arbitrarily large rates.
This ensures that the total LS rate per perturber per unit time
in the reverse direction ($y\rightarrow x$) is also
exactly preserved, again in the limit where
the energy splitting due to fine structure goes to zero.

The standard model was constructed
by reducing the full model, in the following way.
First, the fine structure in all LSJ NIST levels within
$1.52\,\eV$ of the ionisation limit of \ion{N}{I}
($\mathrm{2p^{2}\,4p\,^{4}D^{o}}$ and above),
as well as in the ground level of \ion{N}{II},
were collapsed to LS terms, and the transitions connecting
them were also collapsed,
in the manner described in Sect. 2.3.3 of 
\citet{2017MNRAS.464..264A},
based on the formulae given in
\citet{martin1999atomic}.
Second, all levels within $0.88\,\eV$ of the 
ionisation limit of \ion{N}{I} were grouped into 
two super levels, corresponding to the 
$2S+1=2$ (82 levels) and $4$ (80 levels) spin systems.
The super levels, and corresponding super transitions,
were constructed in an analogous way to how 
all fine structure were collapsed.

We verified that the
standard and full non-LTE model atoms give consistent results,
at least on the \mtd{} model solar atmosphere.
The difference between the abundances inferred
from the full and standard non-LTE model atoms
was less than $0.0001\,\dex$ for the five of the lines
in \tab{tab:n1linelist}.

\section{3D non-LTE effects}
\label{results}

\begin{figure*}
    \begin{center}
        \includegraphics[scale=0.68]{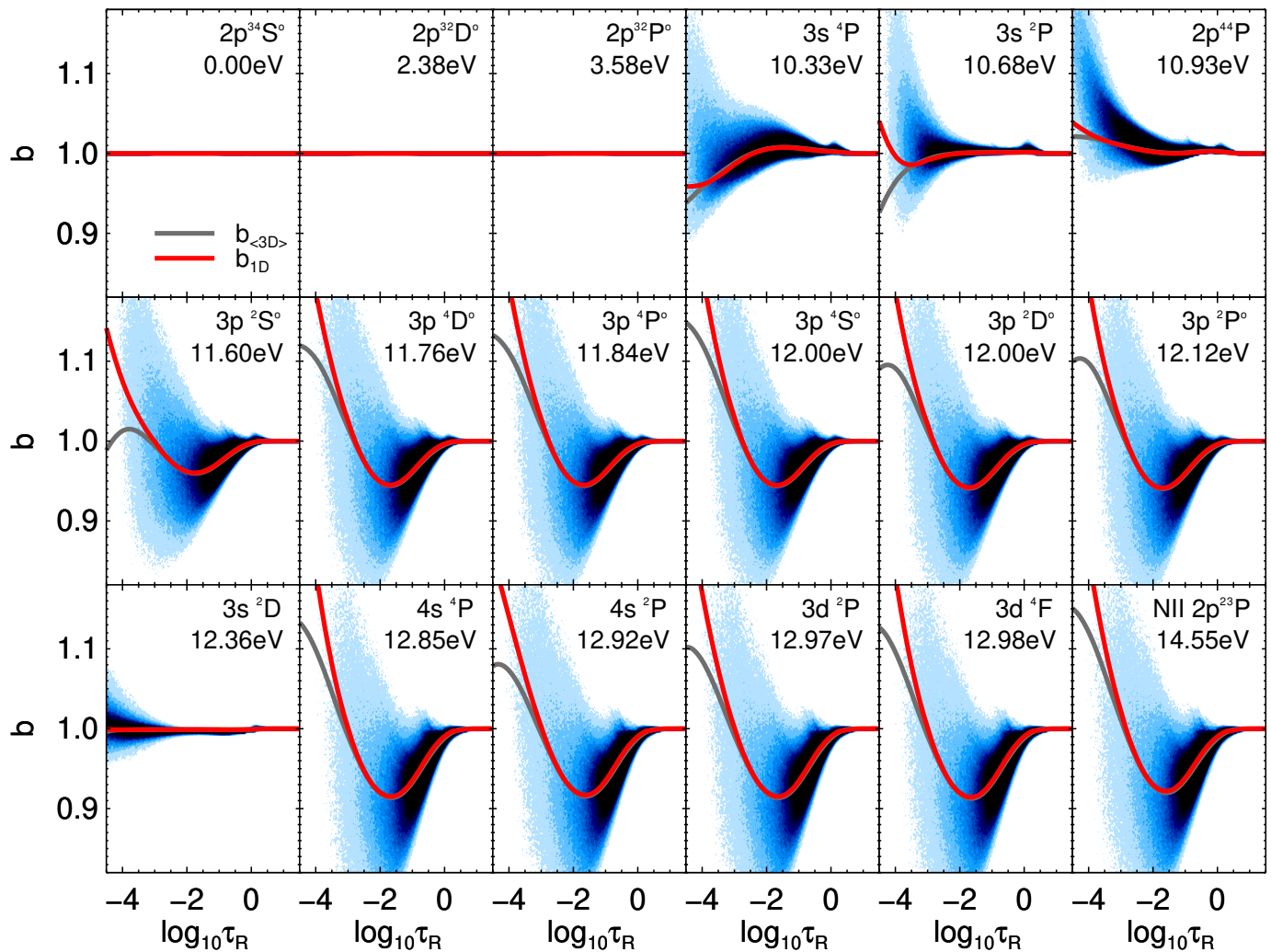}
        \caption{Departure coefficients for the lowest 17 LS terms
        of \ion{N}{I} in order of increasing energy 
        (from left to right, top to bottom),
        as well as the ground level of
        \ion{N}{II} (final panel). 
        The contours show the distributions in the 
        3D model solar atmosphere.  The departure coefficients calculated in the
        \mtd{} and 1D models are overplotted.}
        \label{fig:departure}
    \end{center}
\end{figure*}

\begin{figure*}
    \begin{center}
        \includegraphics[scale=0.31]{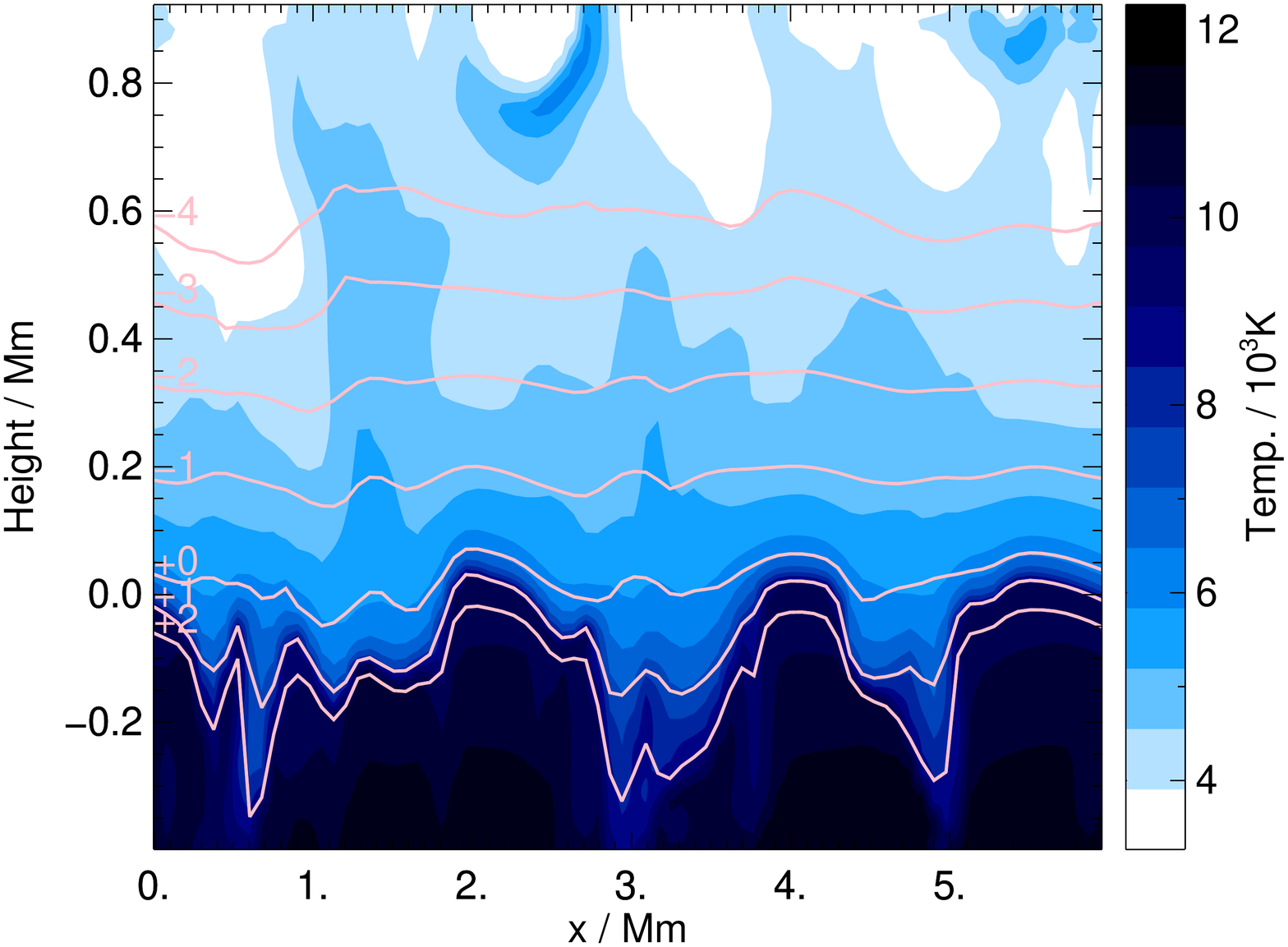}\includegraphics[scale=0.31]{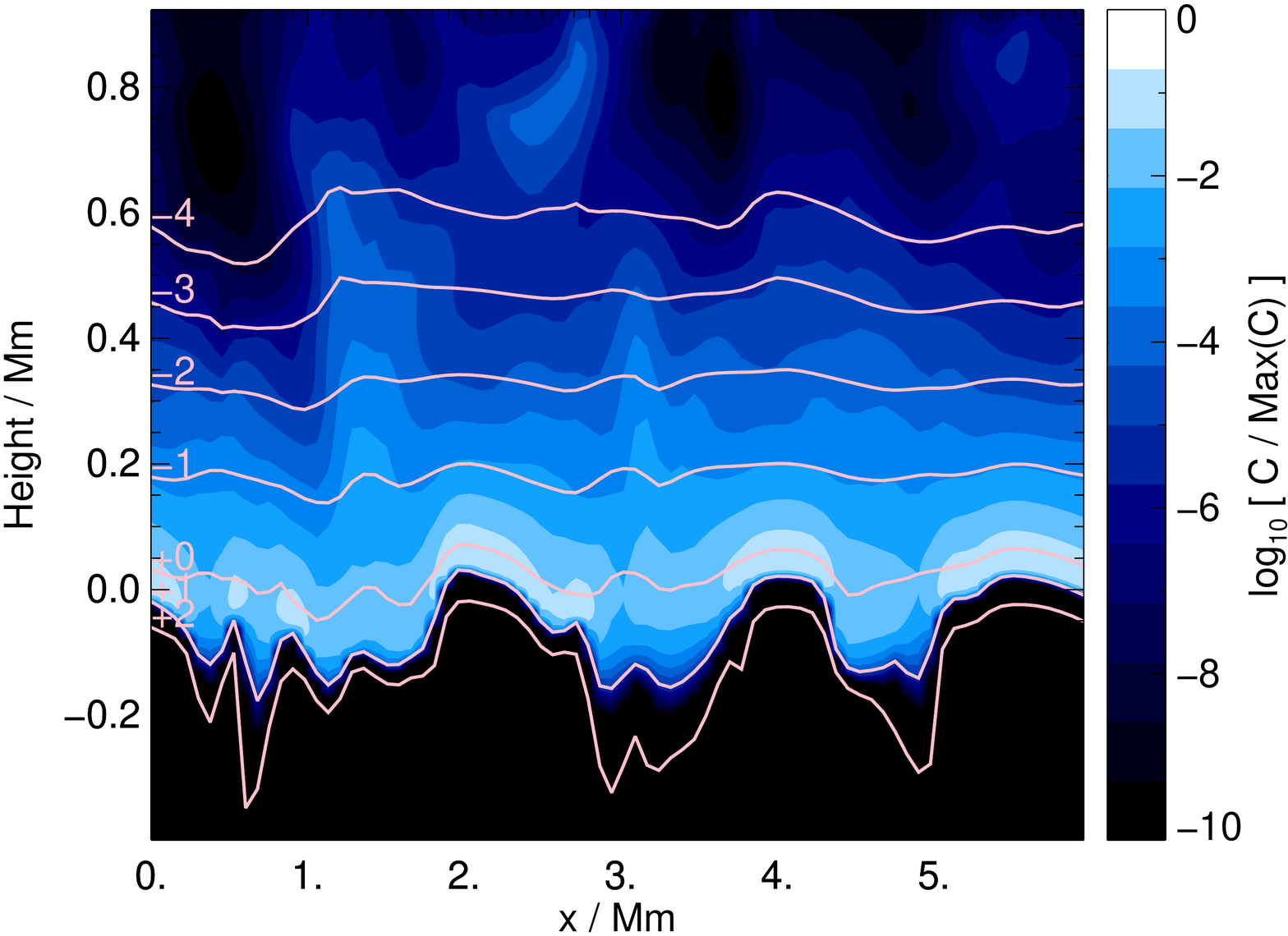}
        \caption{Gas temperature (left) and 
        line forming regions for the
        \ion{N}{I} $868\,\nm$ line (right)
        in a vertical slice of a snapshot of the 3D model solar atmosphere.
        The latter quantity is based on the line-integrated
        contribution function to the line 
        depression in the disk-centre intensity 
        (Eq. 15 of \citealt{2015MNRAS.452.1612A}).
        Contours of constant $\lgr$ are overplotted.}
        \label{fig:cf}
    \end{center}
\end{figure*}

\begin{table*}
\begin{center}
\caption{\ion{N}{I} abundances inferred from different spectrum synthesis models. Results are based on the equivalent widths measured at solar disk centre given in \tab{tab:n1widths}. Unweighted mean abundances $\mu$~and their standard errors $\sigma$~are shown in the final row.}
\label{tab:n1abund}
\begin{tabular}{c | c c c c c c}
\hline
\multirow{2}{*}{Line label} &
\multicolumn{6}{c}{$\lgeps{N}$} \\
& 
\textbf{3D non-LTE} &
3D LTE &
\mtd non-LTE &
\mtd LTE &
1D non-LTE &
1D LTE \\
\hline
\hline
\multicolumn{1}{l|}{\ion{N}{I} 744} &
$\bm{7.767}$ &
$7.779$ &
$7.808$ &
$7.818$ &
$7.798$ &
$7.806$ \\
\multicolumn{1}{l|}{\ion{N}{I} 822} &
$\bm{7.767}$ &
$7.782$ &
$7.810$ &
$7.824$ &
$7.796$ &
$7.806$ \\
\multicolumn{1}{l|}{\ion{N}{I} 863} &
$\bm{7.759}$ &
$7.772$ &
$7.815$ &
$7.826$ &
$7.803$ &
$7.812$ \\
\multicolumn{1}{l|}{\ion{N}{I} 868} &
$\bm{7.768}$ &
$7.784$ &
$7.808$ &
$7.823$ &
$7.794$ &
$7.805$ \\
\multicolumn{1}{l|}{\ion{N}{I} 1011} &
$\bm{7.772}$ &
$7.783$ &
$7.838$ &
$7.848$ &
$7.827$ &
$7.835$ \\
\hline
\multicolumn{1}{c|}{$\mu\pm\sigma$} & 
$\bm{ 7.767\pm 0.002}$ &
\multicolumn{1}{c}{$ 7.780\pm 0.002$} &
\multicolumn{1}{c}{$ 7.816\pm 0.006$} &
\multicolumn{1}{c}{$ 7.828\pm 0.005$} &
\multicolumn{1}{c}{$ 7.804\pm 0.006$} &
\multicolumn{1}{c}{$ 7.813\pm 0.006$} \\
\hline
\end{tabular}
\end{center}
\end{table*}

\subsection{Nature of the non-LTE effect}
\label{resultsnlte}

To understand the nature of the non-LTE effects,
we illustrate the departure coefficients 
for the different \ion{N}{I} levels in \fig{fig:departure}.
The overall picture is qualitatively similar to that
of previous studies
(compare with Fig. 1 of \citealt{2009A&A...498..877C}).
The models indicate that \ion{N}{I} suffers from
photon losses in its lines of high excitation energy:
photons are scattered to large distances in the solar atmosphere
and lost, causing a population cascade to lower energies,
and hence an underpopulation of the most highly excited levels
($\mathrm{3p\,^{2}S^{o}}$, $\chi_{\text{low}}=11.6\,\eV$, and above).
This starts to happen at around $\lgr\approx0.7$,
where the \ion{N}{I} lines, of high excitation energy, begin to form
(\fig{fig:atmos}). This underpopulation is 
typically more severe for levels of higher excitation energy,
as expected from a population cascade.
However, the departure coefficients of the most highly 
excited levels resemble each other, largely owing to 
the efficient hydrogen collisions between them.

Due to number conservation, the
underpopulation of the levels of high excitation
energy must be balanced by a slight overpopulation of the 
levels of intermediate excitation energy:
$\mathrm{3s\,^{4}P}$ ($\chi_{\text{low}}=10.3\,\eV$),
$\mathrm{3s\,^{2}P}$ ($\chi_{\text{low}}=10.7\,\eV$),
and $\mathrm{2p^{4}\,^{4}P}$ ($\chi_{\text{low}}=10.9\,\eV$).
This overpopulation and underpopulation behaviour of
the \ion{N}{I} levels
is comparable to the behaviour of
the levels of high and intermediate excitation energy
respectively in
\ion{C}{I} (Fig. 4 of \citealt{2019A&A...624A.111A}), and of the 
$\mathrm{3p\,^{5}P}$ upper state and
$\mathrm{3s\,^{5}S^{o}}$ lower state 
of the \ion{O}{I} $777\,\nm$ line
(Fig. 4 of \citealt{2018A&A...616A..89A}),
these species also suffering from photon losses.

The $\mathrm{3s\,^{2}D}$ ($12.4\,\eV$) level
is an anomaly in this picture.
Unlike the other high-excitation levels, the
departure coefficients of this level stay relatively close to unity. 
This level has a different core 
($\mathrm{2p^{2}\,^{1}D}$) than the other high-excitation levels in
the figure ($\mathrm{2p^{2}\,^{3}P}$).
As a result, it is only weakly coupled to the rest of the 
\ion{N}{I} system: the electron collisions involving this level
are typically an order of magnitude less efficient than 
collisions involving other levels of comparable energy;
while the hydrogen collisions are neglected from
the non-LTE model, due to Eqs 8 and 9 of \citet{2016A&amp;ARv..24....9B}.

The \ion{N}{I} levels of low excitation energy
$\mathrm{2p^{3}\,^{4}S^{o}}$ ($0\,\eV$),
$\mathrm{2p^{3}\,^{2}D^{o}}$ ($2.4\,\eV$), and
$\mathrm{2p^{3}\,^{2}P^{o}}$ ($3.6\,\eV$) 
retain LTE populations even very high up in the atmosphere:
because of the large energy gap between these levels and
the rest of the \ion{N}{I} system, they are
much more highly populated, and as such their populations
are only minutely perturbed
by the departures from LTE in upper levels.

We present the abundances inferred from the different
spectrum synthesis models 
in \tab{tab:n1abund}.
Taking non-LTE effects into account, 
the \ion{N}{I} lines generally become stronger,
meaning that lower abundances are inferred in 3D non-LTE, 
compared to in 3D LTE.
This can be seen 
in \tab{tab:n1abund}, where the 3D non-LTE $-$ 3D LTE abundance
difference is 
around
$-0.01\,\dex$, averaged over the five \ion{N}{I} lines.

The \ion{N}{I} $744\,\nm$, $822\,\nm$,
$863\,\nm$, and $868\,\nm$ lines,
have the $\mathrm{3s\,^{4}P}$ or $\mathrm{3s\,^{2}P}$
levels of intermediate excitation energy as the lower level,
which slightly overpopulate, and different
levels of high excitation energy as the upper level,
which underpopulate.
They thus suffer from both an enhanced opacity effect
(the line opacity going as the departure coefficient of
the lower level)
and a reduced source function effect
(the line source function going as the ratio of the
departure coefficients of the upper to the lower levels;
\citealt{2003rtsa.book.....R}), both acting to
slightly strengthen the line
(see the integrand of Eq. 15 of \citealt{2015MNRAS.452.1612A}).

The \ion{N}{I} $1011\,\nm$ line is more highly excited.
Both the lower and upper levels underpopulate, 
but as explained above the upper level does so to a greater extent.
Consequently the opacity and source function effects act in opposition;
this line, and indeed the infrared \ion{N}{I} lines in general
(see Table 4 of \citealt{2009A&A...498..877C}),
are thus less sensitive to departures from LTE.
Even so, the source function effect dominates, and the 
\ion{N}{I} $1011\,\nm$ line is stronger in non-LTE,
\tab{tab:n1abund} showing that
the 3D non-LTE $-$ 3D LTE 
abundance difference is $-0.01\,\dex$.

\subsection{Nature of the 3D effect}
\label{results3d}

To aid understanding, it is useful to disentangle the 
3D effect on spectral line formation into two separate effects:
the direct effect,
due to the granulation \citep{2009LRSP....6....2N}
that is present in the 3D model solar atmosphere
but absent in the \mtd{} models,
and the indirect effect, due to differences in the atmospheric 
mean stratification between 
the \mtd{} model, and 1D models.
In the Sun, the \ion{N}{I} lines are susceptible to both
types of 3D effects. 
They work in competition, however, as we explain below.

We investigate the direct 3D effect
by illustrating in \fig{fig:cf} the contribution function
of a typical \ion{N}{I} line in a vertical slice
of a snapshot of the 3D model solar atmosphere.
The \ion{N}{I} lines considered in this study
are all of rather high excitation energy.
They thus preferentially form in high temperature regions:
in the deep atmosphere $\lgr\approx0$,
and in the hot upflowing granules, rather than
in the cool intergranular lanes.
This can be seen in \fig{fig:cf}:
as expected, the contribution function 
is largest (brightest, in the figure) in the deep atmosphere 
around the bumps in the contours of equal
optical depth, which trace the solar granulation.
As also seen for the outer wings of the Balmer lines
of high excitation energy
(see Fig. 2 of \citealt{2018A&A...615A.139A}),
this direct 3D effect acts to strengthen the \ion{N}{I} lines,
due to the granules typically having 
steeper vertical temperature gradients 
than compared to the mean temperature structure.

Although the \ion{N}{I} line formation is biased towards
the granules in the deep atmosphere 
$-0.2\lesssim\lgr\lesssim0.7$,
some additional line formation does occur in the upper atmosphere.
\fig{fig:cf} reveals that even at $\lgr\approx-1$ to $-2$,
minor line formation can occur in the hot temperature
regions above the intergranular lanes
(in the reversed granulation; \citealt{2004A&amp;A...416..333R}),
for example at $x\approx1.5\,\mathrm{Mm}$,
$0.1\lesssim \text{height}\lesssim0.5\,\mathrm{Mm}$.
Moreover, in this particular plot a blob of hot gas
at $x\approx2.5\,\mathrm{Mm}$, 
$\text{height}\approx0.8\,\mathrm{Mm}$ ($\lgr<-4$) is apparent.
Some line formation occurs here as well,
although it is three orders of magnitude less efficient than
the line formation in the solar granulation in the deep atmosphere.

These two direct 3D effects,
namely \ion{N}{I} line formation in the solar granules,
and extended line formation in the upper atmosphere,
both act to strengthen the \ion{N}{I} lines and
thus reduce the abundances inferred from the 3D model.
This can be seen in \tab{tab:n1abund},
where the 3D non-LTE $-$ \mtd{} non-LTE abundance
difference is $-0.05\,\dex$, averaged over 
the five \ion{N}{I} lines.
The direct effect occurs in regions of higher gas temperature,
and consequently the abundance difference is largest for the lines of
higher excitation energy, namely
the \ion{N}{I} $863\,\nm$ 
and $1011\,\nm$ lines (around $-0.06\,\dex$),
compared to the weaker lines ($-0.04\,\dex$).

The nature of the indirect 3D effect can be seen in
\fig{fig:atmos}.  There are differences between
the mean temperature stratification of 
the 3D model (traced by the \mtd{} model)
and that of the 1D model. 
At $\lgr=0.5$ the 1D model is about $150\,\mathrm{K}$ hotter, and at 
$\lgr=-0.5$ the 1D model is about $75\,\mathrm{K}$ cooler.
Thus in the \ion{N}{I} line forming regions 
the \mtd{} model has a much shallower temperature gradient 
than the 1D model.
The indirect 3D effect acts to weaken the \ion{N}{I} lines,
leading to higher abundances inferred in the 3D model
compared to in the 1D model.

The indirect 3D effect is weaker than the direct 3D effect.
This can be seen in \tab{tab:n1abund}, where
the \mtd{} non-LTE $-$ 1D non-LTE abundance
difference is only $+0.01\,\dex$, 
averaged over the five \ion{N}{I} lines.
Consequently the overall 3D effect is dominated
by the direct effect, and
the 3D non-LTE $-$ 1D non-LTE abundance
difference is negative: $-0.04\,\dex$,
again averaged over the five \ion{N}{I} lines.

\subsection{3D/non-LTE coupling}
\label{results3n}

It is interesting to briefly consider how the non-LTE effects
discussed in \sect{resultsnlte}, couple to the 3D effects
discussed in \sect{results3d}.
To quantify this 3D/non-LTE coupling, we compare the 
line-by-line non-LTE versus LTE abundance differences
in \tab{tab:n1abund} to each other.
The [3D non-LTE $-$ 3D LTE] $-$
[1D non-LTE $-$ 1D LTE] abundance difference,
is insignificant: $-0.004\,\dex$,
averaged over the five \ion{N}{I} lines.
This coupling is even smaller
if the indirect 3D effect is omitted, by considering the
[3D non-LTE $-$ 3D LTE] $-$
[\mtd{} non-LTE $-$ \mtd{} LTE] abundance difference: $-0.002\,\dex$,
again averaged over the five \ion{N}{I} lines.

The reason for the small 3D/non-LTE coupling can be seen in
\fig{fig:departure}.  The panels show that, in the 
\ion{N}{I} line forming regions $-0.2\lesssim\lgr\lesssim0.7$,
the departure coefficients calculated for different levels
in the \mtd{} and 1D model solar atmospheres
closely follow each other, and also follow the
median of the corresponding distributions in the 3D model.
They only begin to deviate from each other 
higher up in the atmosphere, at $\lgr\lesssim-3$;
here, the \mtd{} results follow the 3D distribution more closely 
than the 1D results.

Thus, the 3D/non-LTE coupling is not very severe
for \ion{N}{I} in the Sun.
We caution, however, that this result does not necessarily
extend to all late-type stars.
The Sun has a very stratified atmosphere 
compared to other late-type stars,
and the coupling is likely to be larger for stars
with stronger granulation contrast;
namely, for hotter, and more metal-poor stars.

\section{The solar nitrogen abundance}
\label{discussion}

\begin{figure*}
    \begin{center}
        \includegraphics[scale=0.31]{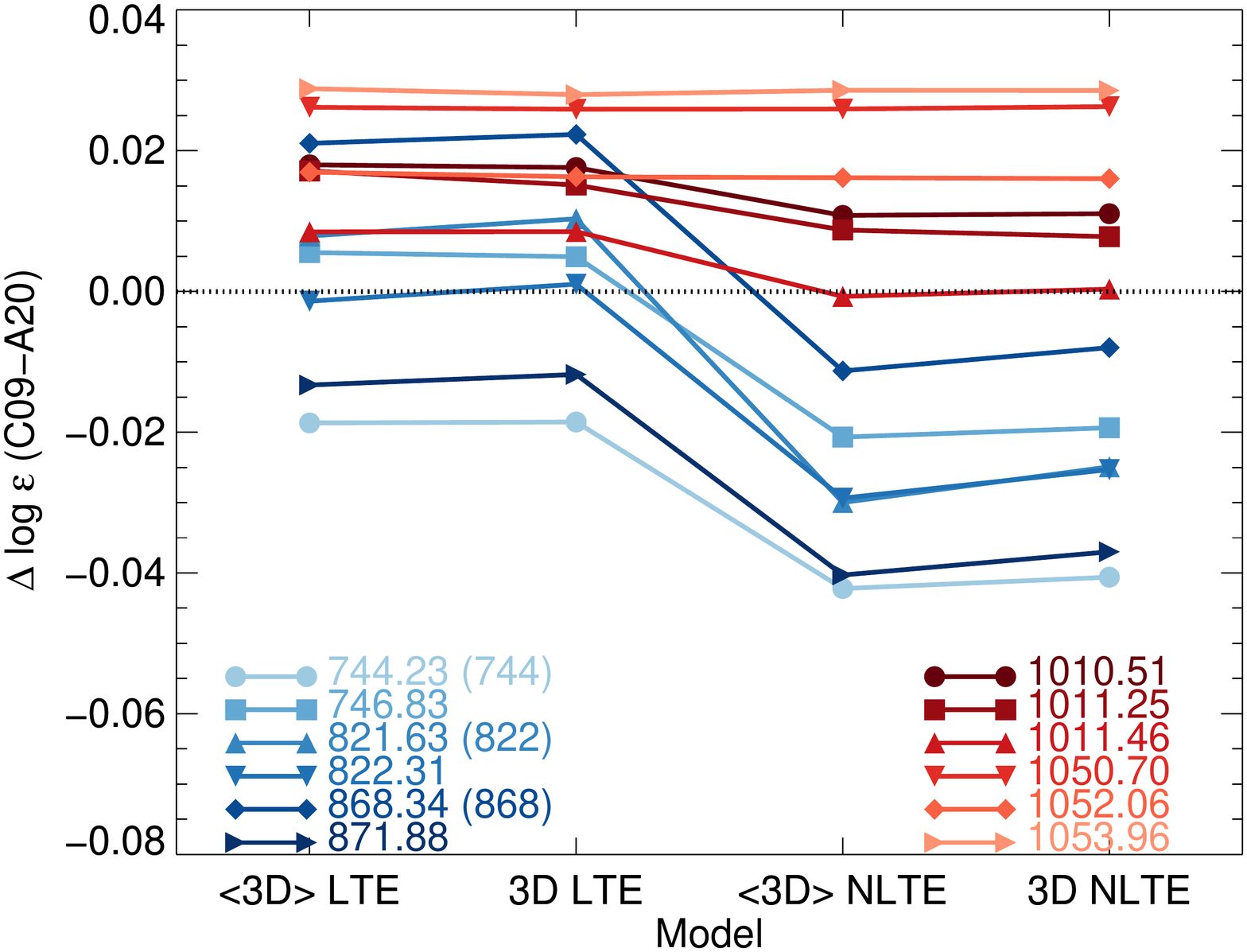}\includegraphics[scale=0.31]{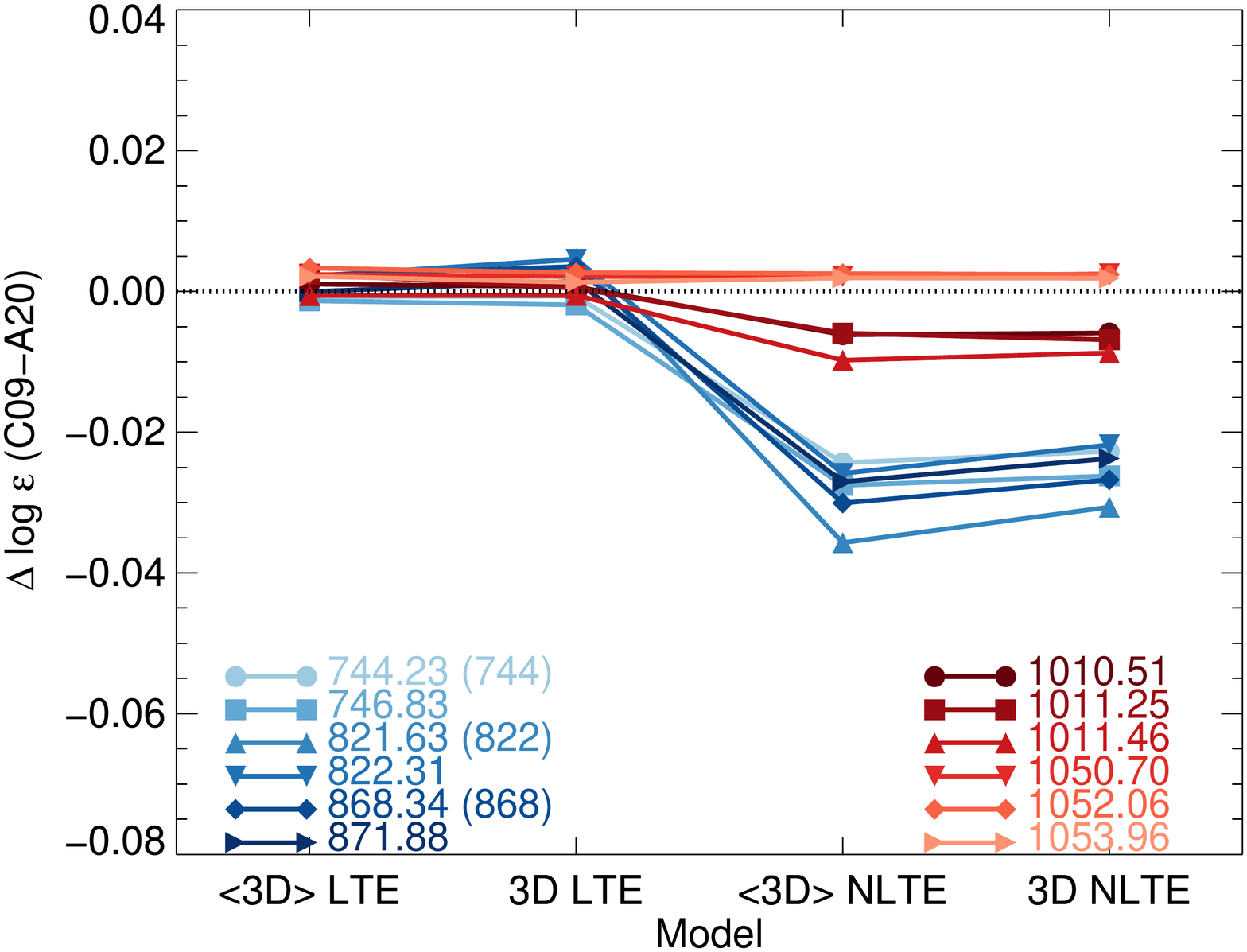}
        \caption{Difference between 
        \citet{2009A&A...498..877C} and this work,
        after adopting the line selection and 
        identical equivalent widths
        \citep{1990A&amp;A...232..225G},
        for different spectrum synthesis models. 
        The three lines in common between that study and the present
        one are marked in parentheses.
        The right panel shows the differences after 
        correcting the abundances for the differences 
        in $\lggf$ between the two studies.}
        \label{fig:compare}
    \end{center}
\end{figure*}

\begin{figure}
    \begin{center}
        \includegraphics[scale=0.31]{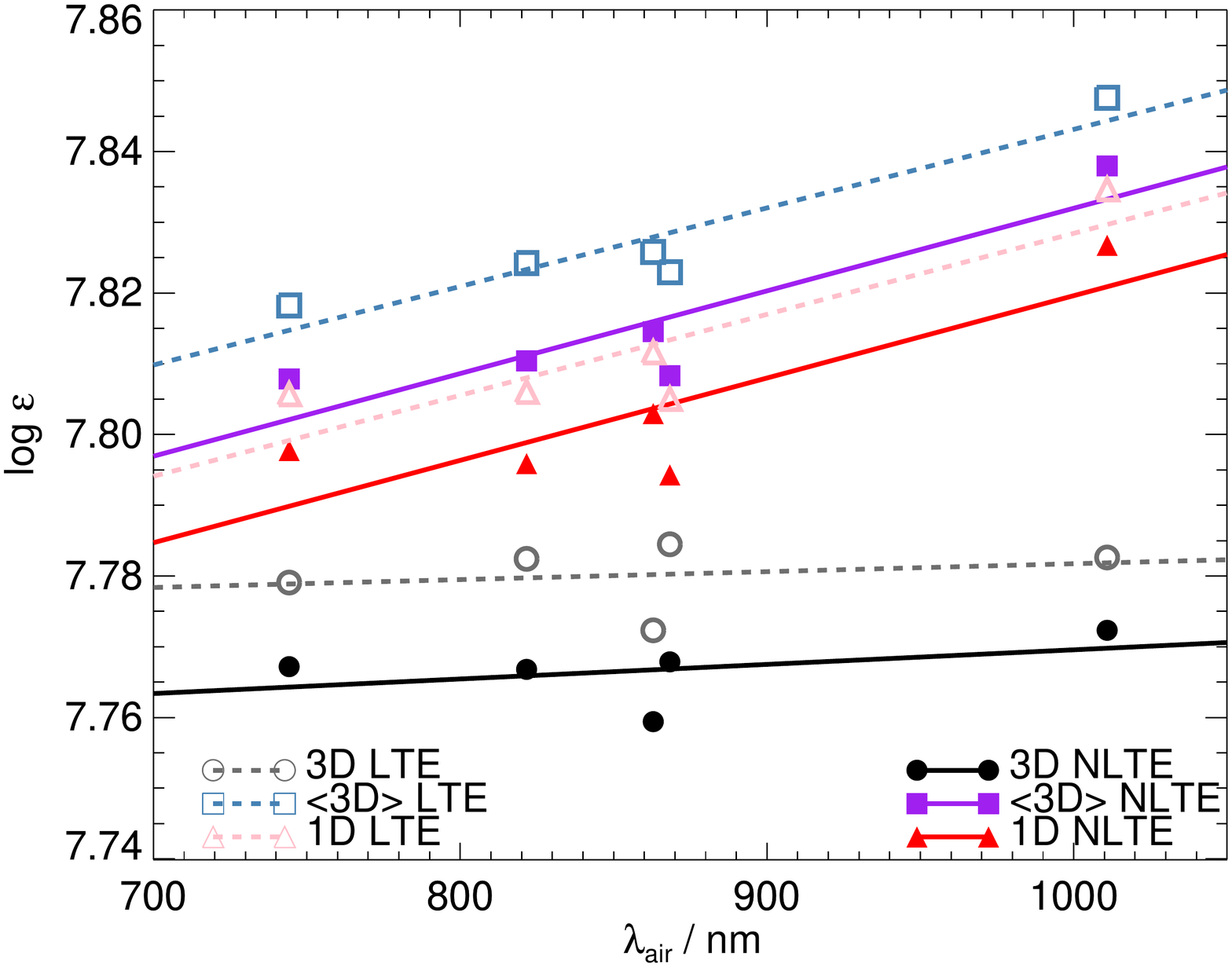}
        \includegraphics[scale=0.31]{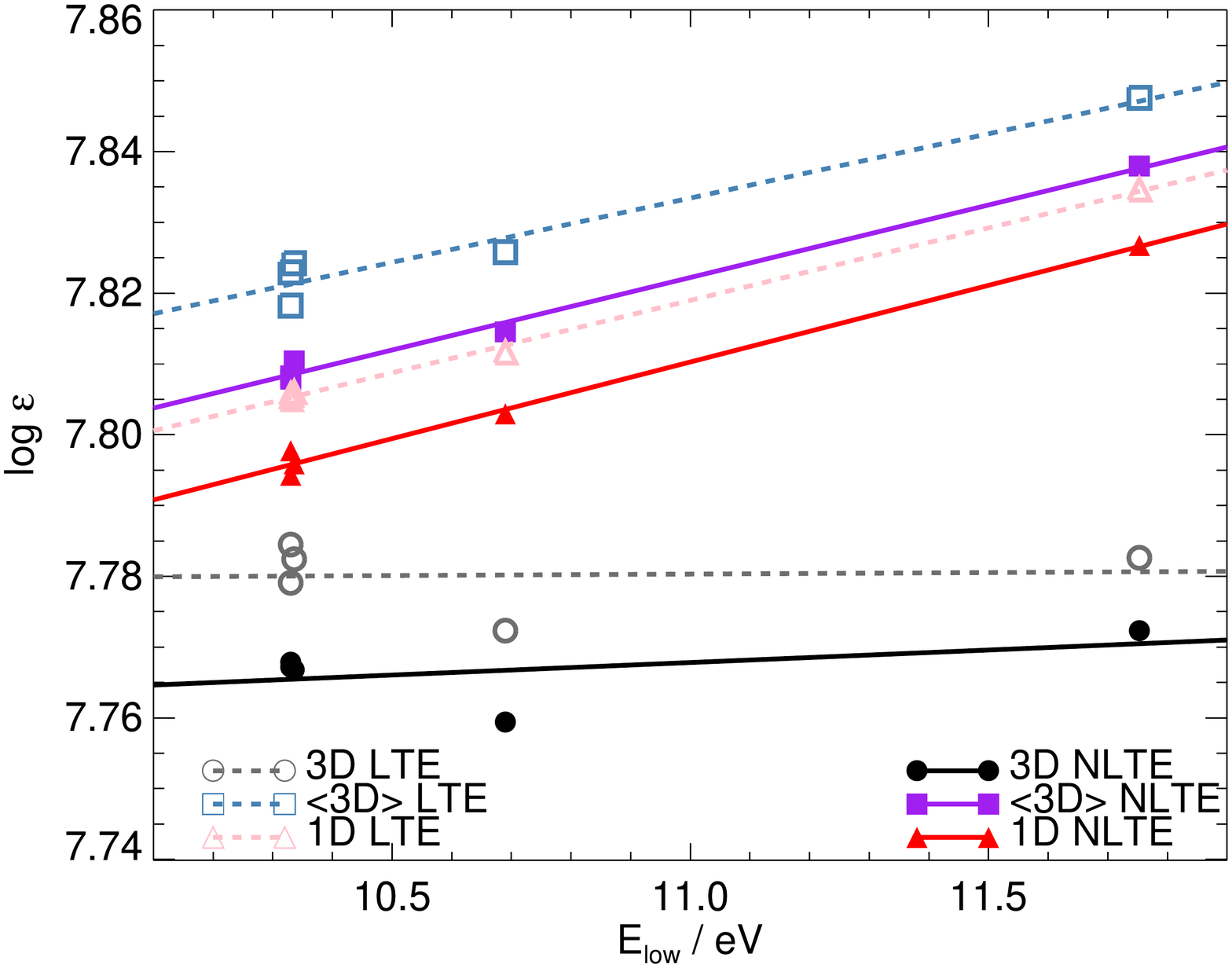}
        \includegraphics[scale=0.31]{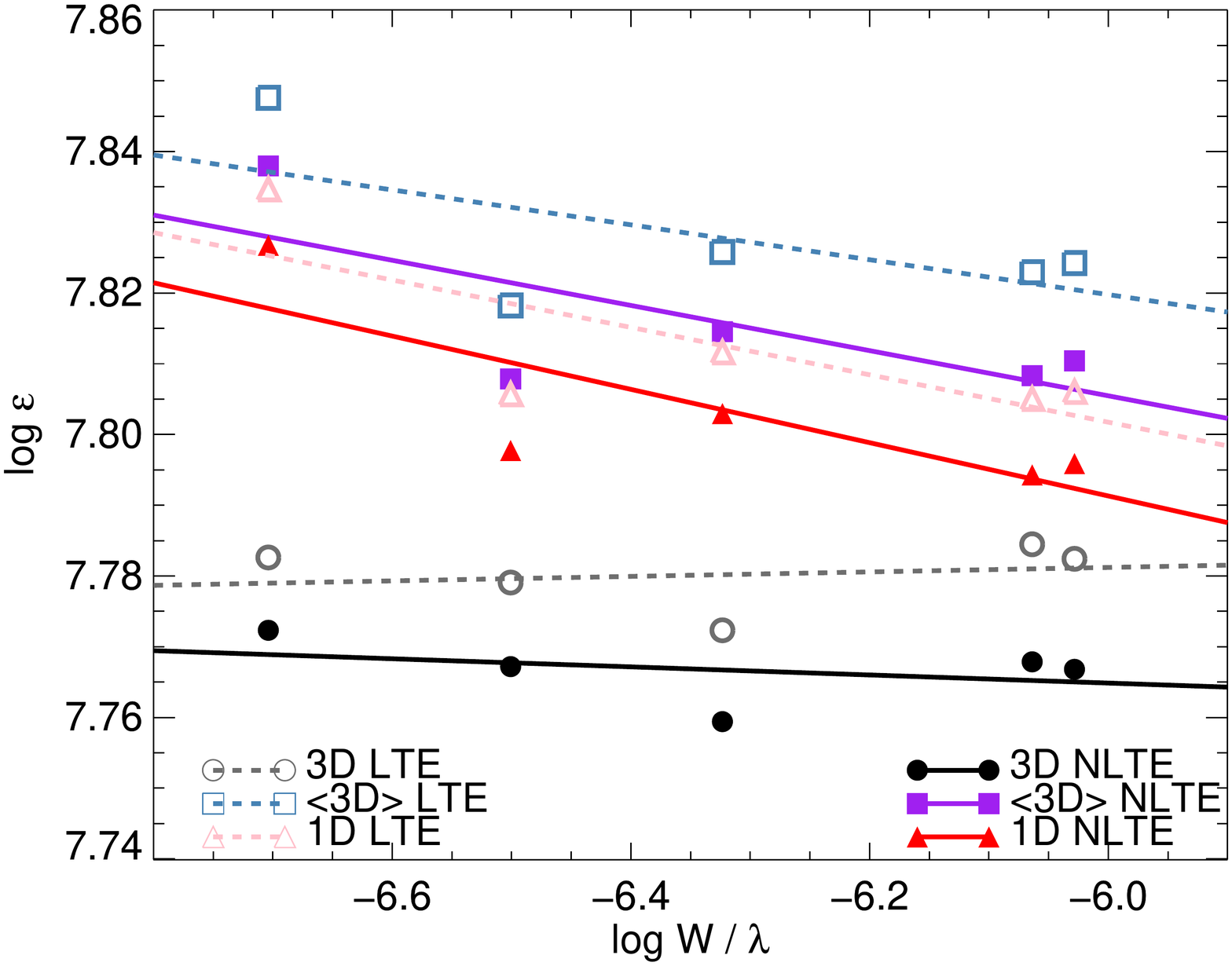}
        \caption{Abundances inferred from the
        five \ion{N}{I} lines in \tab{tab:n1linelist}, using the equivalent
        widths measured at solar disk centre given in
        \tab{tab:n1widths}. Symbols indicate
        results from the different spectrum synthesis models.
        The least squares regressions for each model
        are overplotted.}
        \label{fig:abund}
    \end{center}
\end{figure}

\begin{figure}
    \begin{center}
        \includegraphics[scale=0.31]{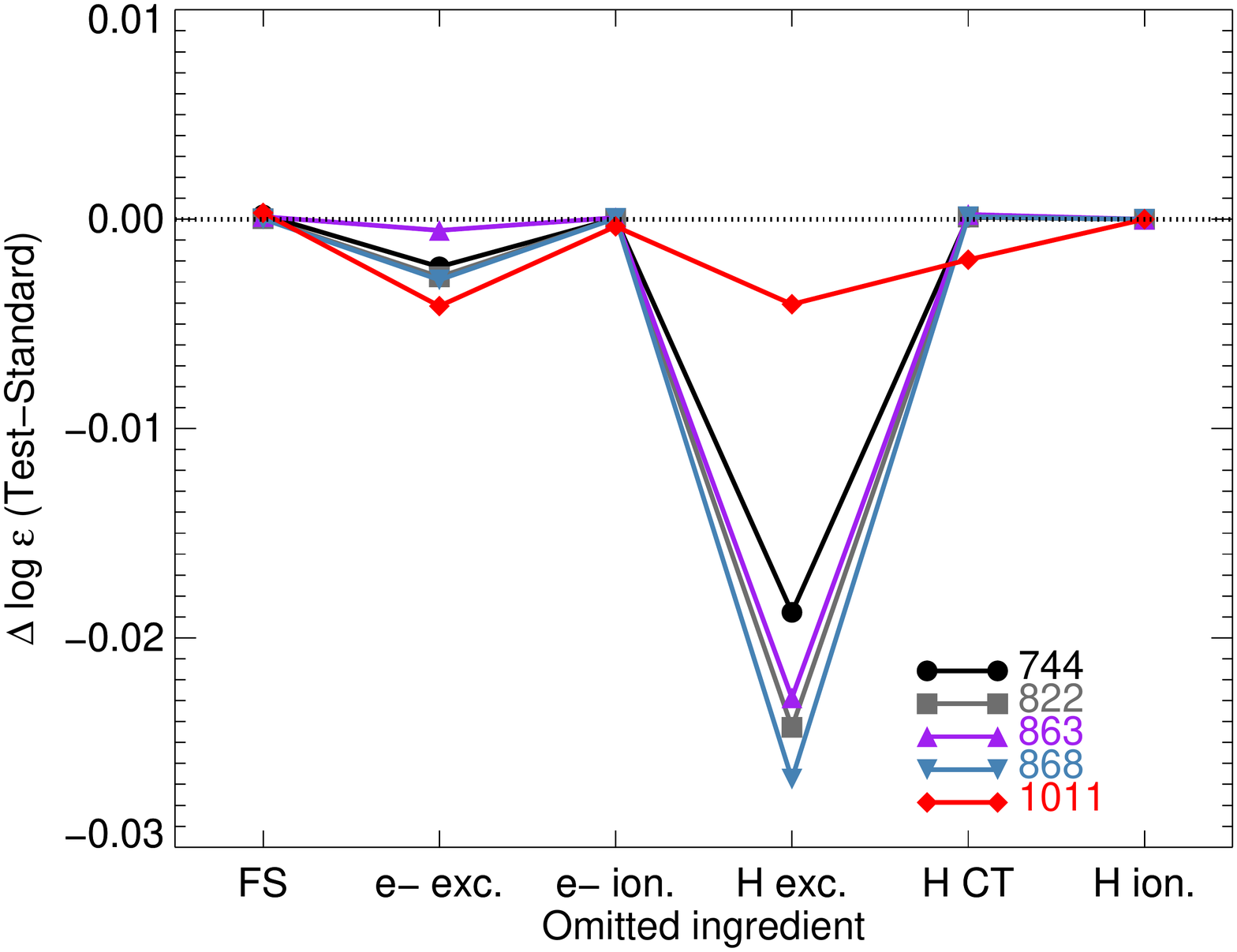}
        \caption{Difference in the inferred abundances
        when switching off the specified
        ingredients in the non-LTE model atom:
        from left to right, fine structure;
        excitation of \ion{N}{I} via electron collisions
        (\eqn{eq:electron1});
        ionisation of \ion{N}{I} via electron collisions
        (\eqn{eq:electron2});
        excitation of \ion{N}{I} via hydrogen collisions
        (\eqn{eq:hydrogen1});
        charge transfer of \ion{N}{I} via hydrogen collisions
        (\eqn{eq:hydrogen2});
        and ionisation of \ion{N}{I} via hydrogen collisions
        (\eqn{eq:hydrogen3}).
        These results
        are based on the standard non-LTE model atom, and
        using the 1D model solar atmosphere.}
        \label{fig:collisions}
    \end{center}
\end{figure}

\subsection{Advocated abundance and uncertainty}
\label{discussionabundances}

Our recommended solar nitrogen abundance from
\ion{N}{I} lines is
$\lgeps{N}=7.77\pm0.05$. This value is determined from the unweighted
mean from the five \ion{N}{I} lines
given by the 3D non-LTE model (\tab{tab:n1abund}),
and is based on fitting the
deblended equivalent widths that were measured at solar disk centre
(\tab{tab:n1widths}).

The 3D non-LTE abundance inferred here is lower than 
that inferred from the 3D LTE model,
and from the \mtd{} and 1D models in both non-LTE 
and LTE.  Both the non-LTE effect 
(\sect{resultsnlte}) and the 3D effect
(\sect{results3d}) strengthen the \ion{N}{I} lines.
Therefore the 3D non-LTE abundance corrections
are negative, relative to the
various \mtd{}, 1D, and LTE models in \tab{tab:n1abund}.

The line-by-line dispersion is low,
corresponding to a standard error of 
$0.002\,\dex$ for the 3D non-LTE model
(\tab{tab:n1abund}).
We thus assume that the overall uncertainty
is dominated by systematics.

To estimate systematic errors
arising from the measurement of the 
deblended equivalent widths of the \ion{N}{I} lines,
the abundance analysis was repeated using 
the older estimates for the equivalent widths given in
\citet{1990A&amp;A...232..225G}.
This same set of equivalent widths 
was used in the abundance analysis of 
\citet{2009A&A...498..877C}.
These equivalent widths are systematically 
somewhat larger: the differences in the inferred abundances
range from $0.02\,\dex$ larger (\ion{N}{I} $868\,\nm$)
to $0.11\,\dex$ larger (\ion{N}{I} $1011\,\nm$);
the mean difference is $0.057\,\dex$.

Following \citet{2015A&amp;A...573A..25S}
we combine, in quadrature, half of this difference 
due to the equivalent widths ($0.028\,\dex$),
with: a) half the difference
between the 3D and \mtd{} results, to quantify 
the uncertainty in the direct 3D effect ($0.025\,\dex$);
b) half the difference between 
the \mtd{} and 1D results, to quantify the uncertainty
in the indirect 3D effect ($0.006\,\dex$);
and c) half the difference between 
the non-LTE and LTE results, to quantify
the uncertainty in the non-LTE effect ($0.007\,\dex$).
We also fold in the uncertainty in the oscillator strengths,
taken to be $0.03\,\dex$ based on
their B+/B rankings given on NIST \citep{kramida2012nist}.
This gives the final result of $0.05\,\dex$.

\subsection{Comparison with \citet{2009A&A...498..877C}}
\label{discussioncaffau}

Our advocated result, 
$\lgeps{N}=7.77\pm0.05$, is somewhat
lower than that presented in \citet{2009A&A...498..877C},
namely $\lgeps{N}=7.86\pm0.12$.
That study and the present one are both based purely on
\ion{N}{I} lines. Both are based on
3D radiative-hydrodynamic model atmospheres and
non-LTE radiative transfer: 
however, the present study adopts 
a full 3D non-LTE approach, whereas 
their study adds non-LTE abundance corrections
from their \mtd{} model solar atmosphere,
to their 3D LTE abundances.
Nevertheless, the 3D/non-LTE 
coupling effect is quite weak for 
\ion{N}{I} in the Sun (\sect{results3n}).
Consequently their
``3D LTE $+$ \mtd{} non-LTE'' approximation should
be in good agreement with the full 3D non-LTE approach.

The main differences between the present study and 
that of \citet{2009A&A...498..877C} are in
the \ion{N}{I} line selection and the adopted equivalent widths.
\citet{2009A&A...498..877C} include 
twelve \ion{N}{I} lines, compared to five in the present study.
The two studies have just three lines in common: the
\ion{N}{I} $744\,\nm$, $822\,\nm$,
and $868\,\nm$ lines.
As we discussed in \sect{methoddiagnostics},
a restrictive line selection was adopted here based on
analysing very carefully line shapes, in order to
minimise the impact of unidentified blends
that tend to skew the inferred
abundances upwards. Moreover, \citet{2009A&A...498..877C} adopt
the de-blended equivalent widths from 
\citet{1990A&amp;A...232..225G};
these values tend to be
systematically larger than those advocated in this study
(\sect{discussionabundances}).
This amounts to an abundance difference of 
$0.06\,\dex$~for the five \ion{N}{I} lines in
\tab{tab:n1linelist}.

We illustrate line-by-line abundance differences 
between the present study and that of 
\citet{2009A&A...498..877C}, in 
the left panel of \fig{fig:compare}.
The plot includes the twelve \ion{N}{I} lines from their study and 
is based on their adopted equivalent widths.
The line parameters adopted in the present study (\sect{methodatom}),
are slightly different to
those adopted in \citet{2009A&A...498..877C}.
In the right panel of the figure
we show the same abundance differences
after correcting for the differences
in the adopted oscillator strengths: $\Delta\lggf=-\Delta\lgeps{N}$.

The right panel of \fig{fig:compare} shows that
the 3D LTE and \mtd{} LTE results from this study 
and from \citet{2009A&A...498..877C}
are in remarkable agreement,
after correcting for differences in the adopted equivalent widths and
line parameters. 
It is clear, however, that there are differences
between the two studies, originating from differences
in the non-LTE modelling.
For the three lines in common between the two studies,
these differences reach almost $-0.03\,\dex$.
The main uncertainty in the non-LTE model atom
is perhaps the hydrogen collisions (\sect{discussiontrends}).
\citet{2009A&A...498..877C} adopt 
the Drawin recipe for these processes,
employing a fudge factor of $\sh=1/3$.
While this was perhaps the best available approach at the time,
it is now known that 
the Drawin recipe does not describe the correct underlying physical
mechanism of these processes \citep{2011A&amp;A...530A..94B}.
The non-LTE model atom presented here uses instead
the LCAO/free electron method.
We have previously argued that this
more accurately represents reality, on the basis of both
physical principles, and empirical results
\citep{2018A&A...616A..89A,2019A&A...624A.111A}.

Adopting the same twelve \ion{N}{I} lines
and using identical oscillator strengths and
equivalent widths as \citet{2009A&A...498..877C},
we obtain $\lgeps{N}=7.87$.
This is in almost perfect agreement with their result,
including a similarly 
large standard deviation in the results 
of $0.11\,\dex$ that suggests
neglected blends and
deficiencies in the adopted equivalent widths.
We conclude that this is
the main reason for the
$0.09\,\dex$ larger abundance inferred by
\citet{2009A&A...498..877C}.
While differences in the non-LTE modelling also play
a small role (of the order $0.03\,\dex$),
differences between the two 3D model solar atmospheres appear not
to contribute significantly to this discrepancy.

\subsection{Model dependence}
\label{discussiontrends}

We illustrate the line-by-line abundances as functions of different
line parameters in \fig{fig:abund},
for the different spectrum synthesis models.
The \mtd{} and 1D models give rise to clear trends
in the inferred abundances with respect to the 
wavelength, excitation energy, and reduced equivalent width.
These trends are largely driven by
the strong direct 3D effect on the \ion{N}{I} lines,
that is due to the solar granulation (\sect{results3d}). 
These trends are also reflected in the relatively
large standard errors in 
\tab{tab:n1abund} from these models.

The flat trends given by the 
3D non-LTE and 3D LTE models in \fig{fig:abund},
and the smaller standard errors given in
\tab{tab:n1abund}, is strong support of the reliability
of the 3D model solar atmosphere used in the present work,
at least in the deep layers
where the \ion{N}{I} lines form,
$-0.2\lesssim\lgr\lesssim0.7$.
This result, combined with previous scrutiny 
of the present 3D model
(Sect. 2.1.4 of \citealt{2018A&A...616A..89A}),
and the excellent agreement between the 3D LTE results
of the present study with the 3D LTE results
of \citet{2009A&A...498..877C} based on
an independent \cobold{} simulation
(when using identical oscillator strengths and
equivalent widths; \sect{discussioncaffau}),
all suggest that uncertainties in the 3D radiative-hydrodynamic
simulations do not have a significant impact on the 
error budget in the present study
(\sect{discussionabundances}).

We test the sensitivity of the inferred abundances to different
ingredients in the non-LTE model atom in \fig{fig:collisions}.
As found for \ion{C}{I} (see Fig. 5 of \citealt{2019A&A...624A.111A}),
the non-LTE results appear to be most sensitive to 
the inelastic collisions with neutral hydrogen
that lead to excitation of \ion{N}{I}.
In the most extreme case,
for the \ion{N}{I} $868\,\nm$ line,
the inferred abundance changes by almost $0.03\,\dex$ 
when hydrogen collisions are switched off;
the mean difference for the five 
\ion{N}{I} lines is $0.02\,\dex$.
It is interesting that 
the \ion{N}{I} $1011\,\nm$ line clearly has 
a lower sensitivity
to the hydrogen collisions than the other four lines:
this likely reflects that the line is the
least sensitive to departures from LTE in general, owing to the
competing non-LTE effects on it (\sect{resultsnlte}).

\fig{fig:collisions} shows that 
the hydrogen collisions play an important role in
the non-LTE modelling. We have previously
motivated our combined LCAO/free electron method employed here,
by studying the centre-to-limb variation
of lines of \ion{O}{I} \citep{2018A&A...616A..89A} 
and \ion{C}{I} \citep{2019A&A...624A.111A}.
If the hydrogen collisions are the largest uncertainty
in the non-LTE model atom, this places an upper bound
of $0.02\,\dex$ on the uncertainty that propagates to the
inferred solar nitrogen abundance.

\section{Conclusion}
\label{conclusion}

We have presented an analysis of the solar photospheric
nitrogen abundance, employing a 
full 3D non-LTE approach, and a model atom that uses
physically-motivated descriptions for the inelastic
collisions between \ion{N}{I} and free electrons,
and between \ion{N}{I} and neutral hydrogen.
We dissected the line formation properties of the \ion{N}{I} lines,
explaining how both non-LTE photon losses, and 
3D granulation effects, both
act to reduce the abundances inferred from
the \ion{N}{I} lines of high excitation energy,
by around $0.01\,\dex$ and $0.04\,\dex$ respectively.
Our advocated value is $\lgeps{N}=7.77\pm0.05$.

There is currently an exceptional interest in revisiting the 
solar elemental composition, because of the
long-standing solar modelling problem.
Standard solar interior models fail to reproduce
several observational constraints precisely determined from
helioseismology \citep[e.g.][]{2008PhR...457..217B}.
It has been noted that replacing the 
standard set of solar abundances of 
\citet{2009ARA&amp;A..47..481A},
with the older canonical compilations 
of \citet{1989GeCoA..53..197A},
\citet{1993oee..conf...15G}, or
\citet{1998SSRv...85..161G},
would help resolve the problem.
However, new analyses suggest that the origin
of the discrepancy is more complex than simply
the adopted solar abundances
\citep{2015Natur.517...56B,2019A&A...621A..33B}.

It is thus important to verify that the 
standard set of solar abundances are indeed robust.
For nitrogen, our advocated value 
of $\lgeps{N}=7.77\pm0.05$ is in excellent agreement
with that presented by
\citet{2009ARA&amp;A..47..481A} from \ion{N}{I} lines,
namely $\lgeps{N}=7.78\pm0.05$.
However, it is slightly lower than
the standard value that they advocate,
namely $\lgeps{N}=7.83\pm0.05$. 
This latter value is larger because it 
folds in results from the NH rotational-vibrational lines
($\Delta\nu=1$), from which \citet{2009ARA&amp;A..47..481A}
obtain $\lgeps{N}=7.88\pm0.03$.
This discrepancy between the atomic and molecular diagnostics
of more than $0.1\,\dex$ is a
cause of some concern.
New molecular data that are available for NH
\citep{2014JChPh.141e4310B,2015JChPh.143b6101B}
and CN \citep{2014ApJS..210...23B}, as well as updates
to the 3D model atmosphere, may well be enough to resolve
the discrepancy. This should be addressed in a future study.

\begin{acknowledgements}
We thank the anonymous referee for their helpful suggestions.
AMA, PSB, and
JG acknowledge support from the Swedish Research Council (VR 2016-03765),
and the project grants ``The New Milky Way'' 
(KAW 2013.0052) 
and ``Probing charge- and mass- transfer reactions on the atomic level''
(KAW 2018.0028) from the Knut and Alice Wallenberg Foundation.
MA gratefully acknowledges funding from
the Australian Research Council
(grants DP150100250 and FL110100012).
Funding for the Stellar Astrophysics Centre is provided by The Danish
National Research Foundation (grant DNRF106).
Some of the computations were performed on resources provided by 
the Swedish National Infrastructure for Computing (SNIC) at 
the Multidisciplinary Center for Advanced Computational Science (UPPMAX) 
and at the High Performance Computing Center North (HPC2N)
under project SNIC 2019/3-532.
This work was supported by computational resources provided by 
the Australian Government through the 
National Computational Infrastructure (NCI)
under the National Computational Merit Allocation Scheme.
\end{acknowledgements}

\bibliographystyle{aa} 
\bibliography{/Users/ama51/Documents/work/papers/bibl.bib}
\label{lastpage}
\end{document}